# Phish Phinder: A Game Design Approach to Enhance User Confidence in Mitigating Phishing Attacks

Gaurav Misra[1], N.A.G. Arachchilage[1] and Shlomo Berkovsky[2]

[1]Australian Centre for Cyber Security
School of Engineering and Information Technology
University of New South Wales, Canberra, Australia
e-mail: {g.misra; nalin.asanka}@adfa.edu.au
[2]Data61, CSIRO
13 Garden St. Eveleigh
NSW 2015 Australia
e-mail: Shlomo.berkovsky@data61.csiro.au

## Abstract

Phishing is an especially challenging cyber security threat as it does not attack computer systems, but targets the user who works on that system by relying on the vulnerability of their decision-making ability. Phishing attacks can be used to gather sensitive information from victims and can have devastating impact if they are successful in deceiving the user. Several anti-phishing tools have been designed and implemented but they have been unable to solve the problem adequately. This failure is often due to security experts overlooking the human element and ignoring their fallibility in making trust decisions online. In this paper, we present Phish Phinder, a serious game designed to enhance the user's confidence in mitigating phishing attacks by providing them with both conceptual and procedural knowledge about phishing. The user is trained through a series of gamified challenges, designed to educate them about important phishing related concepts, through an interactive user interface. Key elements of the game interface were identified through an empirical study with the aim of enhancing user interaction with the game. We also adopted several persuasive design principles while designing Phish Phinder to enhance phishing avoidance behaviour among users.

## 1. Introduction

In March 2016, John Podesta, the chairman of the Hillary Clinton's presidential campaign, received an email which claimed that "someone had his Google password" (Fig. 1). The email provided various details about the login attempt and went on to reassure him that Google had stopped the attempt. It also contained a link luring him to change his password immediately (Fig. 1). It was later discovered that this email was sent by a Russian hacking group that targeted Podesta. The attackers used bit.ly, a web address shortening service (Bitly, 2017), to mask the full URL in the browser. Unfortunately for Podesta, he was deceived into believing that the email was legitimate thereby disclosing his credentials by clicking the malicious link, and causing the leak of Clinton's presidential campaign emails (Krawchenko, 2016). This incident illustrates that humans are susceptible to making poor trust decisions online (Dhamija *et al*, 2006) and are therefore targeted for phishing attacks which can be extremely damaging (Arachchilage *et al*, 2016).

Phishing is an especially challenging cyber security problem as it synthesizes social engineering techniques with technical subterfuge to make the attack successful (Gupta e*t al*, 2017). It has been widely acknowledged as being an important cyber security challenge and there are tools, which have been developed with the aim of preventing it (Sheng *et al*, 2007; Fette *et al*, 2007). However, it has been found that these tools are insufficient as they are unable to detect all phishing attacks (Sheng *et al*, 2007). Moreover, such tools are often designed to warn human users about possible phishing attacks but it is often left to their decision-making capabilities as to whether to download the attachment or click on the link in the email. After all, in most systems, it is impossible to completely circumvent human users (Arachchilage and Love, 2013). In such a scenario, it is imperative to explore methods to educate users and equip them with skills to mitigate such threats effectively. This has led to the shift towards phishing awareness mechanisms which are aimed to train the user to mitigate phishing attacks (Sheng *et al*, 2007; Arachchilage and Love, 2013; Arachchilage *et al*, 2016). A challenging problem, however, has been the difficulty that humans have in making effective and accurate trust decisions when dealing with phishing attacks (Kumaraguru *et al*, 2008; Kirlappos and Sasse, 2012; Arachchilage and Love, 2014). A key aspect, which is not addressed in many previous phishing awareness mechanisms, is the user's self-confidence in dealing with phishing attacks (Kirlappos and Sasse, 2012). It has been found in previous studies that users make better decisions when they are confident about their skills and capability of dealing with the situation (mitigating phishing attacks in this case) (Plant, 1994). Therefore, in this paper, we present *Phish Phinder*, which is a gamified approach to enhance phishing awareness and avoidance behaviour among users, with the following objectives:

a) Integrate self-efficacy into a gamified design for phishing awareness

b) Enhance user interaction and promote phishing threat avoidance behaviour through a series of gamified challenges.

**Fig. 1. The John Podesta emails released by WikiLeaks (Krawchenko, 2016)**

Phish Phinder aims to enhance user confidence in dealing with phishing attacks by providing them with conceptual and procedural knowledge through an interactive user interface. We designed Phish Phinder by adopting a theoretical model underpinning the design and eliciting key interface components by conducting a focus group. We also incorporated several persuasive design principles in Phish Phinder to enhance phishing avoidance behaviour among users.

## 2. Background and Related Work

Automated tools have largely failed to mitigate phishing attacks and even the best anti-phishing tools have been found to miss over 20% of phishing websites (Dhamija *et* al, 2006; Sheng *et al*, 2007; Zhang *et al*, 2007). This failure is compounded by the fact that most systems rely on humans to make sensitive trust decisions during their online activities (Arachchilage *et al*, 2016). This realization has resulted in a shift towards phishing awareness mechanisms (Kirlappos and Sasse, 2012; Arachchilage and Love, 2014) including gamified approaches (Sheng *et al*, 2007; Cone *et al,* 2007) to try and educate users and enhance their threat avoidance behaviour (Foreman, 2004). There are, however, some shortcomings of current gamified approaches which educate users about cyber security challenges. In line with Herley (2009), Kirlappos and Sasse (2012) argue that current security education offers little protection to end users, due to a failure to provide knowledge through an interactive medium rather than an instructional one (Kirlappos and Sasse, 2012). Furthermore, cyber security education needs to prioritize enhancing users' self-confidence in dealing with threats as humans often make better decisions when they are confident in their ability (Plant, 1994).

Phish Phinder addresses these gaps by integrating self-efficacy into a gamified design which is presented to the user through an interactive interface. In the rest of this paper, we discuss the methodology followed to design Phish Phinder and demonstrate an example scenario to illustrate the different user interactions and features of the game.

## 3. Game Design Methodology

The design of Phish Phinder is based on components derived from a theoretical model wrapped up in a story which simplifies the conceptual and procedural knowledge presented to the user by employing persuasive design principles through an interactive interface. It familiarizes players with important concepts related to phishing attacks through challenges presented to them during the game. The challenges in Phish Phinder correspond to the following phishing related concepts, as summarized in previous literature (Fette *et al*, 2007; Zhang *et al*, 2007):

1. *Malicious URLs*: Phishing sites often have URLs which can be spotted and identified as malicious if the user is trained to identify what to look for.

Such URLs may contain IP addresses, for example, which should flag the player's suspicion.
2. ***Lookalike domain***: Phishing attacks often contain URLs or email addresses where some characters are changed to deceive the user. For example, an 'o' may be changed to a '0' and hence an unsuspecting user may be redirected to a malicious site, e.g., www.g0ogle.com.
3. ***Subject line of emails***: A suspicious subject line would contain numerous punctuation marks to attract immediate attention and convey a sense of urgency. They may ask for private details such as password, as in the earlier discussed case with John Podesta, or bank account number claiming the sender to be their bank or their manager at work. The objective would be to train users not to divulge such information by replying to such emails.

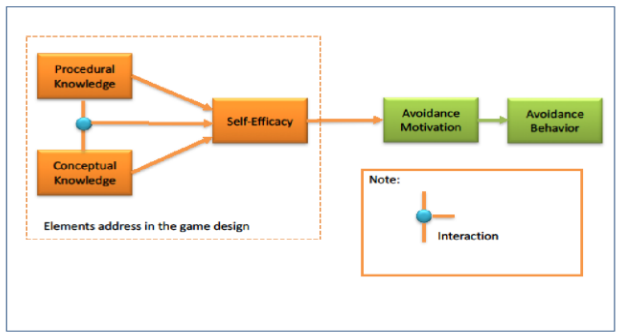

**Fig. 2.  Theoretical model used to design Phish Phinder (Arachchilage and Love, 2013)**

4. ***Display name spoofing***: A common technique used in phishing attacks is that the attacker pretends to be a known individual to the target, e.g., a system administrator, by showing a known name as the sender but using a different email address. The user often does not check the email address of the sender and just responds based on the familiar display name.
5. ***Reply-to spoofing***: Phishing attacks often exploit the fact that users do not check the "reply-to" email address and simply click "reply" and their email is directed to a different email address accessible to the attacker.
6. ***HTML in body of email***: The player would be expected to identify any emails which contain HTML in the body as malicious. While HTML emails have their purpose, the objective of this exercise is to make users aware that such emails can often be malicious and therefore enhance their threat perception about such scenarios.

## 3.1. Theoretical Model

Phish Phinder aims to enhance users' self-confidence in dealing with phishing attacks. This is essential as users are known to make better trust decisions when they are confident in their knowledge and ability to mitigate a threat (Plant, 1994). According to the findings of Arachchilage and Love (2013), computer users' self-efficacy can be enhanced by providing them with both conceptual as well as procedural knowledge. Earlier, McCormick (1997) had also noted that it is essential to provide the user with procedural as well as conceptual knowledge to educate them about a problem adequately. Keeping these observations in mind, Phish Phinder is designed to provide both conceptual as well as procedural knowledge about phishing attacks to users to enhance their self-confidence in dealing with such threats. Enhanced self-efficacy will lead to increased avoidance motivation and ultimately enhance phishing avoidance behaviour among users (Arachchilage and Love, 2013). This goal is achieved by Phish Phinder by deriving elements from a theoretical model described by Arachchilage and Love (2013) (shown in Fig. 2).

In Phish Phinder, the conceptual knowledge is provided to users by explaining why a URL (or email) represents a phishing attack. This conceptual knowledge will be useful for them to identify phishing attacks in the subsequent challenges during the game and applying these previously learned concepts would enhance their procedural knowledge as well. In this way, self-efficacy is integrated into the design of Phish Phinder by employing this theoretical model and wrapping it in a story developed using key persuasive design principles. This story is presented to users through a user-centered interface designed after conducting an empirical investigation through a focus group (described in Section 3.3). This was done to maximize the user interaction with the game as attracting their attention to the problem is an essential first step of any cyber security awareness mechanism (Kirlappos and Sasse, 2012). The scope of the empirical investigation was limited to identification of key elements of the user interface as the knowledge representation in Phish Phinder is based on the theoretical model (Fig. 2) and needed to be preserved and communicated to the user through the story.

### 3.2. The Story

The most important component of Phish Phinder is the story through which the phishing related concepts are presented to the user. We employed several persuasive design principles in Phish Phinder with the aim of enhancing phishing avoidance behaviour (Oinas-Kukkonen and Harjumaa, 2009). The story of Phish Phinder is similar to the one used in "Anti-phishing Phil" (Sheng et al, 2007). We adopted similar characters, which include *Johnny*, a small fish in a pond and the character essayed by the player. The pond is also inhabited by a big fish, *Shifu*, a more experienced and knowledgeable fish, and randomly generated *worms*, each of which represents a URL or an email. Phish Phinder is, however, different to "Anti-phishing Phil" (Sheng et al, 2007) as it integrates self-efficacy in the game design to enhance phishing avoidance motivation and behaviour among users.

Johnny's objective is to eat worms to become a big fish (like Shifu) but he should be careful and avoid eating bad worms. Each worm is randomly generated during the game and is associated with either a URL or an email which is shown to the player. Thus, the player is provided with several classification challenges, identifying whether a worm represents a phishing URL (or email) or a legitimate one, which is the "primary task support" of this game, in accordance with persuasive design requirements (Oinas-Kukkonen and Harjumaa, 2009). The player gets 100 points for a correct classification (eating a good worm or avoiding a malicious worm). If the player makes an erroneous classification, he/she loses one of the 5 lives available at each level of the game.

Phish Phinder is structured in several levels and the narrative builds up as a metaphor of Johnny's life. He eats worms and keeps learning more about phishing to progress through the levels with the aim of eventually becoming as big and knowledgeable as Shifu. Such visible targets are an effective way to motivate players during gameplay (Oinas-Kukkonen and Harjumaa, 2009). The levels are also designed to motivate the player to learn concepts with an increasing level of complexity (e.g. combining different types of phishing concepts into one worm in higher levels of the game). The player has a limited amount of time to complete all challenges in each level (e.g. 10 minutes for level 1) and move to the higher level. The repetition of conceptual examination (through each randomly selected worm) combined with a learning curve, offered by the increasing levels of difficulty, enhances the procedural knowledge of the player and enables them to apply the knowledge they have gained during the game. In this way, adding to the users' conceptual knowledge (through each challenge) as well as their procedural knowledge (through increasing levels and repetition of conceptual examination) would positively impact their self-efficacy to mitigate phishing attacks (Arachchilage and Love, 2013).

If the player is unsure whether a worm is malicious or not, he/she can ask Shifu for guidance. Shifu will then advise about the phishing related concept specific to that worm. The advice provided by Shifu is shown to the player in messages such as, "*a company name followed by a hyphen in a URL is generally a scam*" or "*website addresses associated with numbers in the front are generally scams*" or "*your bank will not send an email to ask you about your account number*", which were crafted by security experts and phrased in a simple and easy-to-comprehend manner. Whenever the player solicits Shifu's help, 60 seconds would be taken off the game clock as a safeguarding cost. Shifu's guidance (when the player is confused) and encouragement (when the player makes a right decision) during gameplay are essential for dialogue support which is an important component of persuasive systems (Oinas-Kukkonen and Harjumaa, 2009).

**3.3. Focus Group: Eliciting Key Components of the Interface Design**

With the aim of creating a user-centered and interactive interface for Phish Phinder, we conducted a focus group to identify key components which needed to be addressed while designing the user interface. The aim of conducting this empirical investigation was to ensure that we included the users' preferences during the design process to maximize user interaction with the game. After all, users will only learn about phishing concepts through the game if they are able to enjoy playing it and maximizing interaction is essential to enhance awareness (Kirlappos and Sasse, 2012).

We invited 6 individuals from our university (4 postgraduate students and 2 academics), between the ages of 26 and 34, to share their opinions about how the game should be designed. All the participants had been using smartphones for at least 4 years. The participants were given a description of the story before arriving for the discussion. They were encouraged to voice their opinion about how they visualized the design and flow of the game. The moderator asked specific questions about the look and feel of the game to elicit ideas for representing the characters of the game as well as the training provided to the players. After interacting with the participants during the focus group, we identified the following key components that needed to be incorporated into the game design:

1. *Narrative* – The participants were unanimous in their opinion that the game, while being educational and aiming to spread awareness of phishing, should have an evolving story which would keep the player engaged. One participant, for example, said "*I only play games which take me somewhere.*

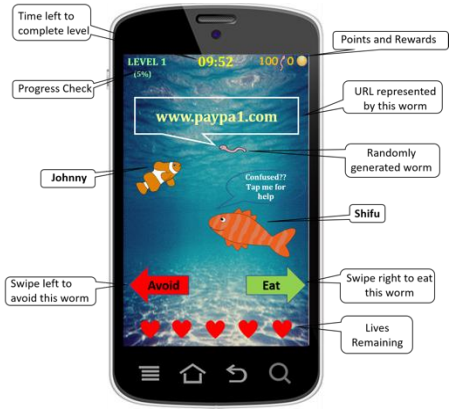

**Fig. 3. Various components of Phish Phinder represented through the user interface**

*I don't like to play games which do not evolve with time.*" Another participant mentioned that he "*would not waste time in a game if the landscape didn't change*" referring to the desired evolution of the narrative of the game.
2. *Rewards* - The game should incorporate a rewards mechanism to incentivise players and keep their engagement levels high by rewarding them for their performance in the game. This has also been established in previous evaluations of gamified approaches (Michallef and Arachchilage, 2017). One participant in our focus group mentioned an instance when he "*did not mind paying money to purchase in-game rewards to speed up the progress*". In Phish Phinder, rewards would only be available by completing levels and learning about phishing related concepts.
3. *Progress Check* – The player should be able view their progress and look back at the training they have received. Moreover, the participants opined that the progress of the player should be contextualized in relation to the remaining game challenges to convince them that successfully completing the game (and thus completing the training) is an achievable target. This can be done relative to the storyline of the game and "*showing the main character's journey would be important*" according to the participants.
4. **Seamless Interaction** – The participants felt that all communication with the player should happen within the gaming environment and should look like *"a part of the scenery"*. The participants suggested that the player should not feel he is navigating away from the game when feedback is provided or when he is making a classification decision (eat/avoid).

The user interface of Phish Phinder, incorporating all the elements identified through the empirical investigation, is shown in Fig 3. The player can "swipe left" to avoid a worm if he/she thinks it represents a malicious URL/email or "swipe right" to make Johnny eat it if he/she thinks it represents a legitimate URL/email. The player can receive rewards in the form of medals (shown in Fig. 3) on completing bonus challenges. When such a bonus worm (having a different colour) comes along, the player will have a limited amount of time (e.g. 30 seconds), depending on the level of the game, to make the decision whether to eat it or not. These challenges are intended to test whether the players can apply learnt concepts under time pressure. Moreover, such bonuses and rewards are known to motivate players to remain engaged with games (Fogg, 2002; Oinas-Kukkonen and Harjumaa, 2009).

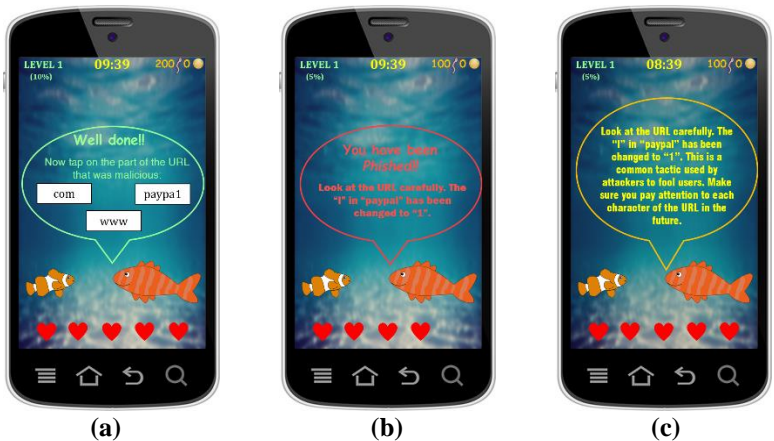

**Fig. 4.** Response corresponding to a situation where the player (a) Avoided the worm, (b) Ate the worm, (c) Asked Shifu for help

## 4. Phish Phinder – User Interaction

To better illustrate the user interaction in Phish Phinder, let us consider the example scenario shown in Fig. 3. The URL shown to the player in Fig. 3, "*www.paypa1.com*" is malicious as the letter "*l*" in "*paypal*" has been changed to the number "*1*". It falls under "malicious URLs" in the types of challenges mentioned in

Section 3. The scenarios corresponding to the three possible actions are (shown in Fig. 4):

- a) *Avoid* the worm by swiping left if he thinks that the URL is malicious. This is the <u>right decision</u> and Shifu will congratulate him and ask him to specify what led him to this conclusion, hence testing the player's conceptual knowledge as it is important that the player not only identifies phishing attacks but also learns how to do so (Fig. 4a).
- b) *Eat* the worm by swiping right if he thinks that the URL is legitimate. This is the <u>wrong choice</u> and Shifu provides him with feedback regarding why he was wrong and which part of the URL represented a phishing attack (the 'l' in "paypal" has been changed to a '1'). The player will also lose a life in this case as shown in Fig. 4b.
- c) *Ask Shifu* for <u>help</u> if he is unsure about the URL who will provide with feedback about the particular URL and the nature of the phishing attack (Fig. 4c). 60 seconds are taken off the game clock as a safeguarding cost.

## 5. Conclusion and Future Work

This paper focuses on designing an innovative gamified approach that aims to educate users about phishing related concepts and enhance their avoidance behaviour. We presented Phish Phinder, which relies on elements derived from a theoretical model (Arachchilage and Love, 2013) to integrate self-efficacy into a gamified approach to educate users about phishing threats. It has an interactive user interface designed by identifying key components through an empirical investigation and incorporating persuasive design principles to enhance phishing avoidance behaviour.

In future work, we intend to implement Phish Phinder as a mobile application to be used by real users. A detailed analysis of the current spectrum of phishing attacks, outlining how they have evolved, will be conducted to ensure relevant and timely education and awareness is provided through Phish Phinder. We plan to conduct user studies to empirically investigate how the knowledge (the interaction effect of both conceptual and procedural) conveyed through the game impacts the users' self-efficacy and ultimately enhances their phishing threat avoidance behaviour. We will also explore the memorability of phishing related concepts provided through Phish Phinder by conducting longitudinal studies with users. Such empirical evaluation will inform us about the usability and effectiveness of Phish Phinder.

Finally, we believe that the design principles described in this paper can be adapted to tackle other important cyber security challenges, such as online authentication, password management and data privacy. This has the potential to enhance avoidance behaviour among users and promote good practices relevant to these threats.